\documentclass[aps,pra,twocolumn]{revtex4}
\usepackage{graphicx,epsfig}
\usepackage{times}
\usepackage{amsmath,amssymb,wasysym}
\usepackage{dsfont} 
\usepackage{color}
\usepackage[linktocpage=true,colorlinks=true, pdfborder={0 0 0},linkcolor=blue,citecolor=blue,urlcolor=blue,anchorcolor=black,linkcolor=blue]{hyperref}

\newcommand{\op}[1]{\hat{#1}}

\newcommand{\ident}{\mathds{1}}

\newcommand{\rem}[1]{}

\newcommand{\imagc}[1]{\text{Im}\,#1}

\newcommand{\real}[1]{\text{Re}(#1)}

\newcommand{\realc}[1]{\text{Re}\,#1}

\newcommand{\bra}[1]{\langle#1|}
\newcommand{\ket}[1]{|#1\rangle}
\newcommand{\transpose}{{\mbox{\footnotesize T}}}

\newcommand{\braket}[2]{\langle#1|#2\rangle}

\newcommand{\sandwich}[3]{\langle#1|#2|#3\rangle}

\newcommand{\quantumaverage}[1]{\langle #1 \rangle}
\newcommand{\ensembleaverage}[1]{{\overline{#1}}}
\DeclareMathOperator{\trace}{Tr}
\DeclareMathOperator{\sprod}{\circ}
\newcommand{\Hs}{\op{H}_0}
\newcommand{\Hp}{\op{H}_1}
\newcommand{\Ha}{\op{H}}
\newcommand{\Htot}{\op{H}_{\text{tot}}}
\newcommand{\Hnoise}{\op{H}_{\text{noise}}}
\newcommand{\Hnoisej}{\op{H}_{\text{noise},j}}
\newcommand{\Hnoisea}{\op{H}_{\text{noise},1}}
\newcommand{\Heff}{\op{H}_{\text{eff}}}
\newcommand{\oL}{{\cal{L}}}
\newcommand{\oLs}{{\cal{L}}_0}
\newcommand{\oLp}{{\cal{L}}_1}
\newcommand{\oLnoise}{{\cal{L}}_{\text{noise}}}
\newcommand{\oLnoisej}{{\cal{L}}_{\text{noise},j}}
\newcommand{\noiseps}{\gamma}
\newcommand{\pf}{\omega}
\newcommand{\order}{n}
\newcommand{\nsources}{K}
\newcommand{\kappac}{\kappa_\text{c}}
\newcommand{\PT}{${\cal{PT}}$}

\begin{document}

\title{Robustness of exceptional-point-based sensors against parametric noise: The role of Hamiltonian and Liouvillian degeneracies}
\author{Jan Wiersig}
\affiliation{Institut f{\"u}r Physik, Otto-von-Guericke-Universit{\"a}t Magdeburg, Postfach 4120, D-39016 Magdeburg, Germany}
\email{jan.wiersig@ovgu.de}
\date{\today}
\begin{abstract}
Recent experiments have demonstrated the feasibility of exploiting spectral singularities in open quantum and wave systems, so-called exceptional points, for sensors with strongly enhanced response. Here, we study theoretically the influence of classical parametric noise on the performance of such sensors. Within a Lindblad-type formalism for stochastic Hamiltonians we discuss the resolvability of frequency splittings and the dynamical stability of the sensor, and show that these properties are interrelated. Of central importance are the different features of exceptional points in the spectra of the Hamiltonian and the corresponding Liouvillian. Two realistic examples, a parity-time-symmetric dimer and a whispering-gallery microcavity with asymmetric backscattering, illustrate the findings.  
\end{abstract}
\maketitle

\section{Introduction}
\label{sec:intro}
The physics of exceptional points (EPs) has become an extensively studied field. In the mathematical literature, they have been introduced already in the sixties of the past century by Kato as degeneracies of order $\order$ of a non-Hermitian linear operator where not only $\order$ eigenvalues but also the corresponding eigenstates coalesce~\cite{Kato66}. That these EPs may give rise to interesting effects in open quantum and wave systems has been proposed in the beginning of this century~\cite{Heiss00,Berry04,Heiss04}. 
The existence of EPs in real physical systems has been first demonstrated in experiments on microwave cavities~\cite{DGH01,DDG04,DFM07}. Later followed experiments on optical microcavities~\cite{LYM09,ZOHY10,POL14,POL16,RZZ19}, coupled atom-cavity systems~\cite{CKL10}, photonic lattices~\cite{RBM12}, nonuniformly pumped lasers~\cite{RotterNC14}, exciton-polariton billiards~\cite{GEB15}, ultrasonic cavities~\cite{SKM16}, high-dielectric spheres in the microwave regime~\cite{WHL19}. 

When a non-Hermitian Hamiltonian $\Hs$ at an EP of order~$\order$ (short: EP$_\order$) is subjected to a perturbation of strength~$\varepsilon$,
\begin{equation}\label{eq:H}
\op{H} = \Hs+\varepsilon\Hp \ ,
\end{equation}
then the resulting energy (or frequency) splittings are generically proportional to the $\order$th root of $\varepsilon$, a fact that was central to the definition of EPs by Kato~\cite{Kato66}. This scaling has to be contrasted with the linear scaling in the case of degeneracies in Hermitian Hamiltonians, so-called diabolic points (DP). 
It has been suggested to exploit these larger splittings at EPs for sensing applications~\cite{Wiersig14b}. This has been studied theoretically in the context of microcavity sensors for single-particle detection~\cite{Wiersig14b,Wiersig16,KYH18,KW19}, parity-time (\PT)-symmetric coupled optical cavities~\cite{HHH15}, coupled nanobeam cavities~\cite{ZYZ15}, (\PT-symmetric) ring laser gyroscopes~\cite{Sunada17,RHH17}, \PT-symmetric electronic circuits~\cite{XLK19}, ultra-high Terahertz sensing~\cite{JTZ18}, and for detecting dark matter candidates~\cite{GMT18}, non-Newtonian effects in gravitation~\cite{LCZ19}, and gravitational waves~\cite{LCH20}. Sensing with families of EPs, so-called exceptional surfaces, has been suggested in Ref.~\cite{ZRK19}.

The feasibility of EP-based sensors has been demonstrated experimentally first for single-particle detection using an EP$_2$~\cite{COZ17} and thermal sensing using an EP$_3$~\cite{HHW17}. Subsequently, the concept has been applied to thermal mapping~\cite{ZCZ18} and sensor telemetry~\cite{CSH18}. Implantable microsensors using a wireless system locked to an EP$_2$ have been implemented in Ref.~\cite{DLY19}. A passive wireless sensing system at an EP$_3$ has also been studied~\cite{ZSL19}.
Ring laser gyroscopes with EP-based enhancement have been fabricated and investigated in Refs.~\cite{LLS19,HSC19}. Very recently, EP-based sensors have been successfully applied in plasmonics~\cite{PNC20}.

The impact of quantum noise on EP-based sensors has been studied including external noise~\cite{LC18,Langbein18,ZSH19,CJL19}, arising from quantum fluctuations in the input signal, and internal noise~\cite{LC18,ZSH19,CJL19} associated with the non-Hermiticity of the sensor. 
References~\cite{LC18,Langbein18,CJL19} have claimed that EPs exhibit no advantage in the deep quantum regime, except if nonreciprocity is utilized~\cite{LC18}. In contrast, Ref.~\cite{ZSH19} has demonstrated that an enhanced sensitivity at an EP is possible in the quantum regime. Note that all the above papers considered exclusively additive noise.

EPs subjected to static disorder~\cite{MGK18} and classical temporal noise~\cite{KMH16,WTM19,XLK19} have been also studied. Static disorder simply blurs the spectral features~\cite{MGK18}. In Ref.~\cite{KMH16} it has been shown that the combination of additive classical noise and nonlinear saturation effects give rise to unexpected behavior in \PT-symmetric mechanical systems. 
\PT-symmetric systems exhibit extra noise because of the gain. However, it was demonstrated theoretically for electronic circuits that the extra noise can be made negligibly small~\cite{XLK19}.
Reference~\cite{WTM19} has addressed the impact of classical fluctuations of the system Hamiltonian~$\Hs$, i.e., parametric noise which on the level of a Schr\"odinger equation appears as multiplicative noise. It has been shown that in such a case a \PT-symmetric EP-based sensor can be dynamically unstable if nonlinear effects, which counteract the instability, are ignored. 

The aim of the present paper is to more deeply study the impact of a fluctuating system Hamiltonian~$\Hs$ on the performance of EP-based sensors. In contrast to Ref.~\cite{WTM19} we consider white Gaussian noise which allows us (i) to adapt a transparent Lindblad-type formalism derived from It\^{o} calculus, (ii) to study the resolvability of the frequency splitting and the dynamical stability in a unified manner, and (iii) to relate our findings to the different properties of the degeneracies in the spectrum of the Hamiltonian and a corresponding Liouville operator. The Liouville spectrum has attracted strong interest in recent years as it signals phase transitions in driven dissipative systems~\cite{MBB18}. The Liouville spectrum has also been studied in the context of EPs~\cite{MMC19,Hatano19,AMM20}.

The outline of the paper is as follows. Section~\ref{sec:derivationL} introduces the formalism for non-Hermitian Hamiltonians under the influence of classical white noise. The spectrum of the resulting Liouville operator and its implications on the dynamical stability of the sensor are discussed in Sec.~\ref{sec:spectrumLindblad}. Section~\ref{sec:examples} provides two examples for an illustration of the general theory. Finally, some conclusions are given in Sec.~\ref{sec:conclusion}.

\section{The Lindblad-type master equation for a noisy non-Hermitian Hamiltonian}
\label{sec:derivationL}
In this section we derive a Lindblad-type master equation by extending the approach in Ref.~\cite{NP99} to non-Hermitian systems with a monochromatic pump and several noise sources.
We consider quantum and classical wave systems where the dynamics of the state vector or wave function $\ket{\psi}$ is governed by a Schr\"odinger-type equation ($\hbar$ is set to unity)
\begin{equation}\label{eq:dynamics}
i\frac{d}{dt}\ket{\psi} = \Htot(t)\ket{\psi} \ .
\end{equation}
For optical systems such an equation of motion can be justified in the slowly-varying envelope approximation in the time domain~\cite{Siegman86}, or by exploiting the similarity between the Schr\"odinger equation and the paraxial wave equation that governs the propagation of light along a waveguide~\cite{RMG10}.

We start with the simple-minded formulation for the total Hamiltonian with parametric noise
\begin{equation}\label{eq:Htot}
\Htot(t) = \Ha+\sum_{j=1}^\nsources\xi_j(t)\Hnoisej
\end{equation}
where $\Ha$ and $\Hnoisej$ are time-independent and non-Hermitian operators. $\nsources$ is the number of statistically independent noise sources.
It is convenient to assume that the units of $\xi_j$ and the matrix elements of $\Ha$ are $\text{s}^{-1}$, which renders the matrix elements of $\Hnoisej$ dimensionless.
The real-valued quantities $\xi_j$ describe pairwise-uncorrelated, stationary white Gaussian noise 
\begin{equation}
\ensembleaverage{\xi_j(t)} = 0
\,,\quad
\ensembleaverage{\xi_i(t)\xi_j(t')} = \noiseps_j\delta_{ij}\delta(t-t') \ ,
\end{equation}
where $i, j = 1,\ldots, \nsources$ and the overline denotes an ensemble average over all possible realizations of the noise. The nonnegative parameters $\noiseps_j$ are measures of the strength of the noise in units of $\text{s}^{-1}$.
The more rigorous formulation is based on the It\^{o} differential of the Wiener processes
\begin{equation}
dW_j(t) = \int_t^{t+dt}\xi_j(t')dt'
\end{equation}
and 
\begin{equation}\label{eq:dWW}
\ensembleaverage{dW_j(t)} = 0
\,,\quad
\ensembleaverage{dW_i(t)dW_j(t)} = \noiseps_j\delta_{ij}dt \ .
\end{equation}

\subsection{Wave function}
We assume that each of the noise sources is physical in the sense that it is a limit of a process with finite correlation time. It is textbook knowledge, see, e.g., Ref.~\cite{Gardiner90}, that in such a case the stochastic differential equation should be formulated as a Stratonovich stochastic equation. Hence we write Eq.~(\ref{eq:dynamics}) as 
\begin{equation}\label{eq:Strato}
i\ket{d\psi} = \Ha\ket{\psi}dt + \sum_j\Hnoisej\ket{\psi}\sprod dW_j + e^{-i\pf t}\ket{P} dt
\end{equation}
where $\sprod$ denotes the Stratonovich product. Note that the parametric noise appear here as multiplicative noise. 
We have also introduced a non-noisy coherent pump with the frequency $\pf$ and strength given by the norm of $\ket{P}$. While in the (classical) optical setting such a pump term is natural and routinely used to describe the excitation of the system via a waveguide, see for example Refs.~\cite{POL16,Sunada18,Langbein18}, it is more difficult to justify it in the quantum case, but it is also used in this context, see for instance Ref.~\cite{DCG18}.

Following the standard recipes used for stochastic differential equations~\cite{Gardiner90} we transform the Stratonovich form in Eq.~(\ref{eq:Strato}) to the equivalent It\^{o} form
\begin{equation}\label{eq:ito}
i\ket{d\psi} = \Heff\ket{\psi}dt + \sum_j\Hnoisej\ket{\psi}dW_j + e^{-i\pf t}\ket{P} dt 
\end{equation}
with the effective Hamiltonian
\begin{equation}\label{eq:Heff}
\Heff =  \Ha - \frac{i}{2}\sum_j\noiseps_j \Hnoisej^2 \ .
\end{equation}
Using Eqs.~(\ref{eq:dWW}) and~(\ref{eq:ito}) we get for the ensemble-averaged wave function $\ensembleaverage{\ket{\psi}}$ the inhomogeneous Schr{\"o}dinger equation
\begin{equation}\label{eq:iSe}
i\frac{d}{dt}\ensembleaverage{\ket{\psi}} = \Heff\ensembleaverage{\ket{\psi}} + e^{-i\pf t}\ket{P} \ .
\end{equation}
The effective Hamiltonian $\Heff$ fully captures the dephasing of the ensemble-averaged wave function induced by the noise. 
If at least one of the eigenvalues of $\Heff$ has a nonnegative imaginary part then for generic initial conditions the solution is unbounded. This includes the case of zero imaginary part because of the possible polynomial growth in time at an EP~\cite{DFM07,Heiss10,WTM19}. This kind of instability is not related to the noise but due to the fact that gain is not compensated by loss. We exclude here this possibility by restricting ourselves on systems where all eigenvalues of~$\Heff$ have a negative imaginary part. 
Note that this can embrace \PT-symmetric system Hamiltonians $\Hs$ as the dephasing introduced by the second term in Eq.~(\ref{eq:Heff}) appears on this level usually as small losses. This happens for example, when all $\Hnoisej$ are Hermitian because then $\sum_j\noiseps_j \Hnoisej^2$ is positive semidefinite; for positive semidefinite decay operators see, e.g.,~\cite{Wiersig19}. In such a case, the noise stabilizes the dynamics of the ensemble-averaged wave function.
The wave function in the long-time limit is then given by the particular solution of Eq.~(\ref{eq:iSe}) 
\begin{equation}\label{eq:particular}
\ensembleaverage{\ket{\psi(t)}} = \op{G}(\pf)e^{-i\pf t}\ket{P}
\end{equation}
with the Green's operator (or the resolvent) of $\Heff$
\begin{equation}
\op{G}(\pf) = \left(\pf\ident-\Heff\right)^{-1} \ ,
\end{equation}
which characterizes the response of the wave function to the pump with the frequency~$\pf$; $\ident$ is the identity operator.

\subsection{Density operator}
In most of the physically relevant situations it is not the amplitude, $\ket{\psi}$, that is measured but the intensity, $\braket{\psi}{\psi}$, or more generally, expectation values $\quantumaverage{\op{A}} = \sandwich{\psi}{\op{A}}{\psi}$ where $\op{A}$ describes an observable of interest. An ensemble average gives
\begin{equation}\label{eq:At}
\ensembleaverage{\quantumaverage{\op{A}}}(t) = \trace{\left[\op{\rho}(t)\op{A}\right]}
\end{equation}
where the Hermitian and positive semidefinite operator
\begin{equation}
\op{\rho}(t) = \ensembleaverage{\ket{\psi(t)}\bra{\psi(t)}}
\end{equation}
is interpreted as density operator. For the increment
\begin{equation}
d\op{\rho} =  \ensembleaverage{\ket{\psi+d\psi}\bra{\psi+d\psi}} - \ensembleaverage{\ket{\psi}\bra{\psi}}
\end{equation}
it is straightforward to show with Eqs.~(\ref{eq:dWW}) and~(\ref{eq:ito})
\begin{eqnarray}
\nonumber
d\op{\rho} & = &  -i\left(\Heff\op{\rho} - \op{\rho}\Heff^\dagger\right) dt + \sum_j\noiseps_j\Hnoisej\op{\rho}\Hnoisej^\dagger dt\\
\label{eq:drho}
&& -i\left(\ket{P}\bra{\ensembleaverage{\psi}}e^{-i\pf t} - e^{i\pf t}\ket{\ensembleaverage{\psi}}\bra{P}\right) dt
\end{eqnarray}
where terms $\propto (dt)^2$ have been ignored and $\dagger$ denotes the Hermitian conjugate. Inserting Eq.~(\ref{eq:particular}) into Eq.~(\ref{eq:drho}), we finally get an inhomogeneous Lindblad-type master equation
\begin{equation}\label{eq:drhodt}
\frac{d\op{\rho}}{dt} = \oL\op{\rho} + \op{P}(\pf)
\end{equation}
with the superoperator~$\oL$ acting on the density operator~$\op{\rho}$
\begin{eqnarray}\label{eq:Lindblad}
\oL\op{\rho} & = & -i\left(\Heff\op{\rho}-\op{\rho}\Heff^\dagger\right) + \sum_j\noiseps_j\Hnoisej\op{\rho}\Hnoisej^\dagger 
\end{eqnarray}
and the Hermitian pump operator
\begin{equation}\label{eq:pump}
\op{P}(\pf) = -i\left(\ket{P}\bra{P}\op{G}^\dagger(\pf) - \op{G}(\pf)\ket{P}\bra{P}\right) \ .
\end{equation}
Note that the master equation~(\ref{eq:drhodt}) with Eq.~(\ref{eq:pump}) describes only the long-time dynamics accurately because Eq.~(\ref{eq:particular}) is only valid in this regime. The restriction to the long-time dynamics is completely adequate for our purpose.

We mention that a pump with amplitude noise (independent of the other noise sources) could be included in Eq.~(\ref{eq:Strato}) as well. This would give an additional frequency-independent term $\ket{\tilde{P}}\bra{\tilde{P}}$ in the pump operator~(\ref{eq:pump}) which we do not consider here. A noisy pump frequency $\pf$ can be indirectly included by noisy diagonal elements in the Hamiltonian~(\ref{eq:Htot}). 

In the special case $\nsources = 1$, $\op{P} = 0$, $\Ha = \Ha^\dagger$, and $\Hnoisea = \Hnoisea^\dagger$ Eqs.~(\ref{eq:drhodt}) and~(\ref{eq:Lindblad}) reduce to the Lindblad-type master equation in Ref.~\cite{NP99} (for further references where classical white noise results in a Lindblad-type master equation consult~\cite{MGC16,CBC17}). In this case, Eq.~(\ref{eq:Lindblad}) conserves the trace of $\op{\rho}$ under the dynamics and $\oL$ has a similar structure as in the conventional quantum Lindblad master equation (see e.g.~\cite{BP02}) where the first term describes coherent processes and dissipation and the second term describes quantum jumps. 
In our case, however, the trace of $\op{\rho}$ is not conserved because both $\Ha$ and $\Hnoisej$ may not be Hermitian. In fact, $\Ha$ must be non-Hermitian in order to possess an EP. In a full quantum treatment, one should include the quantum jump terms related to the non-Hermiticity of $\Ha$ and $\Hnoisej$ as well. We do not consider this since we have systems in mind with essentially classical behavior.

In the r.h.s. of Eq.~(\ref{eq:drhodt}), only the second term depends on the pump frequency~$\pf$. The pump operator in Eq.~(\ref{eq:pump}) is proportional to the pump rate squared and to the local density of states summed over $\ket{P}$. Since here the Green's operator of the effective Hamiltonian $\Heff$ enters, we conclude that the relevant frequencies for resonant excitation are given by the eigenvalues of $\Heff$. 

The resolvability of frequency splittings at and near an EP is therefore determined by $\Heff$. 
From the definition of the effective Hamiltonian~(\ref{eq:Heff}) it is clear that the position of an EP in parameter space is here not influenced by the noise if $[\Hs,\Hnoisej^2] = 0$ for $j = 1,\ldots,\nsources$. This might appear as a special condition, but we will see later that it is valid for realistic scenarios.

The first term on the r.h.s. of Eq.~(\ref{eq:drhodt}) determines the stability of the system in terms of bounded intensities. 
To see this it is convenient to move to the Liouville space as we do in the next section.

\section{The Liouville spectrum and stability}
\label{sec:spectrumLindblad}
For an Hilbert space of dimension $N$ the Liouville space is of dimension $N^2$. In Liouville space, $\op{\rho}$ and $\op{P}$ are represented as $N^2$-dimensional vectors and $\oL$ is represented as a $N^2\times N^2$ matrix. The latter is called Liouville operator or short Liouvillian. 
The Liouvillian is a non-Hermitian matrix which describes the time evolution of the density operator $\op{\rho}$. Because of the Hermiticity of $\op{\rho}$ for all times, the Liouvillian has to fulfill a \PT-like symmetry~\cite{HJW10}. This implies that its eigenvalues are either real or come in complex conjugate pairs. 
The degeneracies of the considered Hamiltonian in general differ from the degeneracies of the corresponding Liouvillian~\cite{MMC19}. Consequently, we distinguish in the following between the Hamiltonian EP (HEP), Liouvillian EP (LEP), Hamiltonian DP (HDP), and Liouvillian DP (LDP). 

As discussed in the previous section, the stability of the system in terms of bounded intensities is determined by the first term on the r.h.s. of Eq.~(\ref{eq:drhodt}). In Liouville space this stability can be analyzed in terms of the eigenvalues~$\lambda_l$ of the Liouvillian~$\oL$. A stationary state $\op{\rho}$ is in general only possible if all eigenvalues have a nonpositive real part. If one (or more) of the eigenvalues have a positive real part then the components of the density operators diverge in the long-time limit. This, in principle, could be avoided by choosing proper initial conditions and $\op{P}(\pf)$ inside a subspace orthogonal to the corresponding eigenvector(s). This is extremely difficult if not even impossible for a whole interval of values of~$\pf$ which is needed for determining a spectrum.


\subsection{Degeneracies}
Let us first look at the system alone ($\Hp = 0 = \Hnoisej$) with the Liouvillian 
\begin{equation}\label{eq:L0}
\oLs\op{\rho} = -i\left(\Hs\op{\rho}-\op{\rho}\Hs^\dagger\right) \ .
\end{equation}
Consider a system Hamiltonian $\Hs$ at a HDP of order $\order$. We restrict ourselves henceforth on the corresponding subspace of Hilbert space with dimension $\order$. The eigenvalues of $\Hs$ are all equal, $E_j = E$, and the eigenstates $\ket{E_j}$ with $j = 1,\ldots,\order$ can be chosen such that they form an orthogonal basis in the $\order$-dimensional Hilbert space. Importantly, the dyads $\ket{E_i}\bra{E_j}$ are linearly independent right eigenvectors of $\oLs$ for $i, j = 1,\ldots,\order$ with all eigenvalues being equal to $-\Gamma$ with $\Gamma = -2\imagc{E}$. Hence, the Liouville operator is at an LDP of order $\order^2$.

Next, consider the system Hamiltonian $\Hs$ to be at an HEP of order $\order$. $\Hs$ has only one eigenstate $\ket{E_j}$. So, we can construct only one eigenvector of $\oLs$ by using a dyad $\ket{E_j}\bra{E_j}$. Does it mean that the Liouville operator is at an LEP of order $\order^2$? No, not necessarily, because there might be other, nondyadic eigenvectors of $\oLs$. 
This can be shown explicitly for an HEP$_2$ by making use of the Jordan canonical form of the system Hamiltonian
\begin{equation}
\Hs = \left(\begin{array}{cc}
E & 1\\
0 & E\\
\end{array}\right) \ .
\end{equation}
A straightforward calculation shows that the Liouvillian~(\ref{eq:L0}) is given by 
\begin{equation}\label{eq:LEP2}
\oLs = \left(\begin{array}{cccc}
-\Gamma & -i      & i       &  0      \\
0       & -\Gamma & 0       &  i      \\
0       & 0       & -\Gamma & -i      \\
0       & 0       & 0       &  -\Gamma\\
\end{array}\right) 
\end{equation}
where the density operator is represented by the Liouville-space vector $(\rho_{11},\rho_{21},\rho_{12},\rho_{22})^\transpose$ with the superscript $\transpose$ marking the transpose of the vector. The elements are $\rho_{ij} = \ensembleaverage{\psi_i\psi_j^*}$ with the Hilbert-space vector $\ket{\psi} = (\psi_1,\psi_2)^\transpose$. The Liouvillian in Eq.~(\ref{eq:LEP2}) has two right eigenvectors: $(1,0,0,0)^\transpose$ and $(0,1,1,0)^\transpose$. Both have eigenvalue $-\Gamma$. It can be verified that the first eigenvector belongs to an LEP$_3$ and that the second one is independent of this LEP. 
The fact that a Hamiltonian $\Hs$ at an HEP$_2$ implies a LEP of third order for a Liouvillian of the form as in Eq.~(\ref{eq:L0}) has been already observed by direct calculation of an example in Ref.~\cite{AMM20} and is also consistent with the finding that an LEP$_3$ can be generated by varying just two real parameters due to the \PT-like symmetry of the Liouvillian~\cite{Hatano19}.

A simpler argument which covers also the higher dimensional case $\order > 2$ is here provided by examining the time dynamics. A generic initial condition $\ket{\psi(0)}$ at an HEP$_\order$ evolves according to an exponential function multiplied by a polynomial of degree $\order-1$ in time $t$~\cite{Heiss10}. Hence, the initial condition  $\ket{\psi(0)}\bra{\psi(0)}$ for the Liouvillian in Eq.~(\ref{eq:L0}) evolves according to an exponential function multiplied by a polynomial of degree $2\order-2$ in time~$t$. Consequently, the Liouvillian must be at an EP of order $2\order-1$. In the special case $\order = 2$ the LEP is of third order in agreement with the result above. 

\subsection{Stability}
Now we include $\varepsilon\Hp$ and $\Hnoisej$ with the perturbation strength~$\varepsilon$ and the noise strengths~$\noiseps_j$.  
In the beginning let us ignore the last term in Eq.~(\ref{eq:Lindblad}), which in the quantum Lindblad master equation corresponds to quantum jumps:
\begin{eqnarray}\label{eq:Lindbladwqj}
\oL'\op{\rho} & = & -i\left(\Heff\op{\rho}-\op{\rho}\Heff^\dagger\right) \ .
\end{eqnarray}
For such a special Liouvillian all relevant information is present already in the Hamiltonian~\cite{MMC19}, here the effective Hamiltonian~$\Heff$. We can conclude from Eqs.~(\ref{eq:H}) and~(\ref{eq:Heff}) that generically the splittings of the eigenvalues of the Liouvillian are of order $\varepsilon$ (for all $\noiseps_j = 0$) and $\noiseps_j$ (for $\varepsilon = 0 = \noiseps_{i\neq j}$) for an HDP and correspondingly of order $\varepsilon^{1/\order}$ and $\noiseps_j^{1/\order}$ for an HEP$_\order$. This is not in contradiction to the fact that the order of the LEP is $2\order-1$ because the perturbation by $\varepsilon\Hp$ is nongeneric on the level of Eq.~(\ref{eq:Lindbladwqj}). 

Finally, let us consider the full Liouvillian including the last term in Eq.~(\ref{eq:Lindblad}). In the special case $[\Heff,\Hnoisej] = 0$ for all~$j$ the Liouville spectrum is still determined by the spectrum of $\Heff$. The situation is different for $[\Heff,\Hnoisej] \neq 0$. As discussed above, if $\Hs$ is at an HDP than $\oLs$ is at an LDP. A generic perturbation including the noise [with the last term in Eq.~(\ref{eq:Lindblad})] still gives frequency splittings of order $\varepsilon$ and $\noiseps_j$. 
For $\Hs$ being at an HEP$_\order$, the Liouvillian $\oLs$ is at an LEP$_{2\order-1}$. Hence, a generic perturbation gives splittings with different scaling $\varepsilon^{1/\order}$ (for all~$\noiseps_j = 0$) and $\noiseps_j^{1/(2\order-1)}$ (for $\varepsilon = 0 = \noiseps_{i\neq j}$). 
We conclude that the splittings of the eigenvalues of the Liouvillian at an LEP are very sensitive to the strength of the noise. The sensitivity is stronger the higher~$\order$ is. 
This has implications on the dynamical stability of the sensor. A splitting induced by the noise is proportional to $\noiseps_j^{1/3}$ for an HEP$_2$ which for small noise strength $\noiseps_j$ is parametrically large if compared to $\noiseps_j$ and at least one of the splittings has a positive real part. If this real part is larger than the distance of the eigenvalue of $\oLs$ to the imaginary axis, one of the eigenvalues of $\oL$ cross the axis leading to the instability. The same is true for $\order > 2$.

The discussed instability can be compensated by (i) nonlinear effects such as gain saturation and (ii) by adding uniform dissipation to the sensor. While the former possibility may complicate the interpretation of the measurement data, the latter possibility may reduce the resolvability. This can be a problem for EP-based sensors operating close to or at the real frequency axis, such as \PT-symmetric sensors, as will be revealed in the next section. 

\section{Examples}
\label{sec:examples}
In this section we study two realistic scenarios to illustrate the general theory.

\subsection{\PT-symmetric dimer}
\label{sec:example1}
The first example is a \PT-symmetric dimer with the Hamiltonian
\begin{equation}\label{eq:e1H}
\op{H} = \left(\begin{array}{cc}
\omega_0-i\alpha & g\\
g   & \omega_0+i\alpha\\
\end{array}\right) 
\end{equation}
and real-valued frequency $\omega_0$, gain coefficient $\alpha \geq 0$, and coupling strength $g \geq 0$. One site exhibits loss ($-i\alpha$) and one site exhibits gain ($i\alpha$) of the same amount. The system is invariant under simultaneously performing the parity operation (exchanging the sites) and time reversal (turning loss into gain and vice versa). 
The eigenvalues of the Hamiltonian~(\ref{eq:e1H}) are given by
\begin{equation}\label{eq:e1evH}
E_\pm = \omega_0 \pm \sqrt{g^2-\alpha^2} \ .
\end{equation}
The Hamiltonian possesses an HEP$_2$ at $g = \alpha \neq 0$. Considering the deviation $g = \alpha + \varepsilon$ as perturbation with perturbation strength~$\varepsilon$ divides the Hamiltonian~(\ref{eq:e1H}) into a system part describing the sensor and a perturbation part
\begin{equation}\label{eq:e1Hs}
\Hs = \left(\begin{array}{cc}
\omega_0-i\alpha & \alpha\\
\alpha   & \omega_0+i\alpha\\
\end{array}\right) 
\;\,\text{and}\;\,
\varepsilon\Hp = \left(\begin{array}{cc}
0 & \varepsilon\\
\varepsilon & 0\\
\end{array}\right) \ .
\end{equation}
From Eq.~(\ref{eq:e1evH}) it is easy to see that the frequency splitting goes with $\varepsilon^{1/2}$ for small $|\varepsilon| \ll 2\alpha$. For the HDP ($\alpha = 0$) the splitting is linear in $\varepsilon$.

The above \PT-symmetric dimer has also been studied in the context of noise in Ref.~\cite{WTM19}. As in that reference, we consider one source of noise only, a fluctuating on-site frequency detuning,
\begin{equation}\label{eq:e1Hn}
\Hnoisea = \Hnoise = \left(\begin{array}{cc}
1 & 0\\
0    & -1\\
\end{array}\right) \ .
\end{equation}
Note that our model of the noise (white and Gaussian) is very different from the above reference. 

Since $\Hnoise$ is Hermitian and thus $\gamma\Hnoise^2$ with noise strength $\noiseps = \noiseps_1 \geq 0$ is positive semidefinite, the such described fluctuations introduce losses into the wave function dynamics. To see this more concretely, we plug Eqs.~(\ref{eq:e1Hs}) and~(\ref{eq:e1Hn}) into the effective Hamiltonian~(\ref{eq:Heff}) yielding
\begin{equation}\label{eq:e1Heff}
\Heff = \left(\begin{array}{cc}
\omega_0-i\alpha-i\noiseps/2 & \alpha+\varepsilon\\
\alpha+\varepsilon    & \omega_0+i\alpha-i\noiseps/2 \\
\end{array}\right) \ .
\end{equation}
A comparison to Eq.~(\ref{eq:e1H}) shows that here the noise may introduce a uniform damping which shifts the eigenvalue of the effective Hamiltonian by $-i\noiseps/2$ into the lower complex plane. This renders the resolvability of the frequency splitting slightly more difficult. 
It can be seen from Eq.~(\ref{eq:e1Heff}) but also from the fact that $\Hnoise^2 = \ident$ and therefore $[\Hs,\Hnoise^2] = 0$ that the HEP is located at exactly the same position ($g = \alpha$) in parameter space as in the absence of the noise. This is not the case for the Liouvillian as is shown below.

We address the dynamical stability by first considering the density matrix $\rho_{ij} = \ensembleaverage{\psi_i\psi_j^*}$ as Liouville-space vector $(\rho_{11},\rho_{21},\rho_{12},\rho_{22})^\transpose$ where the Hilbert-space vector $\ket{\psi} = (\psi_1,\psi_2)^\transpose$ is defined in the same basis as the Hamiltonian~(\ref{eq:e1H}). From Eq.~(\ref{eq:Lindblad}) we then derive the Liouvillian 
\begin{equation}\label{eq:eqLtot}
\oL = \oLs+\varepsilon\oLp+\noiseps\oLnoise 
\end{equation}
with the system part
\begin{equation}\label{eq:e1Ls}
\oLs = \left(\begin{array}{cccc}
-2\alpha & -i\alpha & i\alpha   &  0      \\
-i\alpha & 0        & 0         & i\alpha     \\
i\alpha  & 0        & 0         & -i\alpha    \\
0        & i\alpha  & -i\alpha  &  2\alpha\\
\end{array}\right) \ ,
\end{equation}
the perturbation part
\begin{equation}\label{eq:e1Lp}
\oLp = \left(\begin{array}{cccc}
0  & -i & i  &  0      \\
-i & 0  & 0  & i     \\
i  & 0  & 0  & -i    \\
0  & i  & -i &  0      \\
\end{array}\right) \ ,
\end{equation}
and the terms related to the noise
\begin{equation}\label{eq:e1Lnoise}
\oLnoise = \left(\begin{array}{cccc}
 0      &  0       &  0       &   0     \\
 0      & -2       &  0       &   0     \\
 0      &  0       & -2       &   0     \\
 0      &  0       &  0       &   0     \\
\end{array}\right) \ .
\end{equation}
In the presence of noise, the Liouvillian of the unperturbed system ($\varepsilon = 0$) is no longer at the LEP. To see this, we calculate the four eigenvalues yielding $\lambda_3 = -2\noiseps$ and for $\varepsilon = 0$ in lowest order in $\noiseps$ (for $\alpha > 0$)
\begin{equation}\label{eq:e1ev}
\lambda_l = 2\alpha^{2/3}\noiseps^{1/3} e^{i2\pi (l-1)/3}
\end{equation}
with $l=0,1,2$. The cubic-root scaling of the latter three eigenvalues stems from departing from the LEP$_3$ as predicted by the general considerations in Sec.~\ref{sec:spectrumLindblad}. 
The eigenvalue~$\lambda_1$ has the largest real part. We define the critical rate
\begin{equation}\label{eq:e1kappac}
\kappac = \frac{\realc{\lambda_1}}{2} = \alpha^{2/3}\noiseps^{1/3} > 0
\end{equation}
which quantifies the dynamical instability of the sensor. Such a noise-induced parametric instability is well known in the theory of parametric resonance, see, e.g.,~\cite{ZP03}. 
In the HDP case ($\alpha = 0$) all eigenvalues of the Liouvillian have a nonpositive real part and therefore no instability exists for all values of $\noiseps$.

The finding of dynamical instability is consistent with Ref.~\cite{WTM19}. Using the realistic estimates $\alpha = 10^{12}\,\text{s}^{-1}$ and $\noiseps = 10^{-12}\,\text{s}^{-1}$ from that reference for optical \PT-symmetric microcavities gives $2\kappac = 10^4\,\text{s}^{-1}$. The time scale for the instability is therefore around $0.1\,$ms. Interestingly, this is two orders of magnitude longer than the result in Ref.~\cite{WTM19} which is based on the actually much more favorable assumption of having colored noise with zero spectral density at the frequency of the HEP. 

The obvious, but in fact, subtle solution of the stability problem was not discussed in Ref.~\cite{WTM19}: the dynamical instability can be removed by a uniform damping of the sensor, i.e., by adding $-i\kappa\ident$ to the system Hamiltonian~$\Hs$ (and hence $-2\kappa\ident$ to $\oLs$), with the rate~$\kappa$ above the critical rate~(\ref{eq:e1kappac}). Unfortunately, this reduces the resolvability since the linewidths broaden significantly proportional to $\noiseps^{1/3}$ which is parametrically larger than the scaling from static disorder which goes with the square root~\cite{MGK18}. 
This makes, for instance, the resolution estimate for gravitational wave detection using the EP-based microcavity optomechanical sensor in Ref.~\cite{LCH20} too optimistic. Liu and coworkers have assumed that the resulting frequency splitting has to overcome the frequency noise variance~$\noiseps$ and determine the minimal perturbation strength $\varepsilon$ to achieve this. From $g/2\pi = 10^7\,$s$^{-1}$ and $\noiseps \approx 3.3\cdot 10^{-4}\,$s$^{-1}$ the resulting relative spatial strain resolution is roughly $10^{-24}$ which is better than the advanced LIGO (Laser Interferometer Gravitational-Wave Observatory) interferometers. In contrast, our analysis shows that the stabilization requires the frequency splitting to overcome the linewidth~$2\kappac \approx 1.1\cdot 10^4\,$s$^{-1}$ which gives a relative spatial strain resolution of around $10^{-9}$.

Note that the first higher-order correction in $\noiseps$ to the eigenvalues~(\ref{eq:e1ev}) is $-2\noiseps/3$ which stabilizes the dynamics if $\noiseps > 3\kappac$. Hence, an intentional and significant increase of the noise may remove the dynamical instability. But it does not improve the resolvability because of the damping term $-i\noiseps/2$ in the effective Hamiltonian~(\ref{eq:e1Heff}).
We conclude that for EP-based sensors operating at or close to the real frequency axis even small noise can indirectly, via the necessary stabilization, spoil the resolvability considerably. 

\begin{figure}[ht]
\includegraphics[width=0.95\columnwidth]{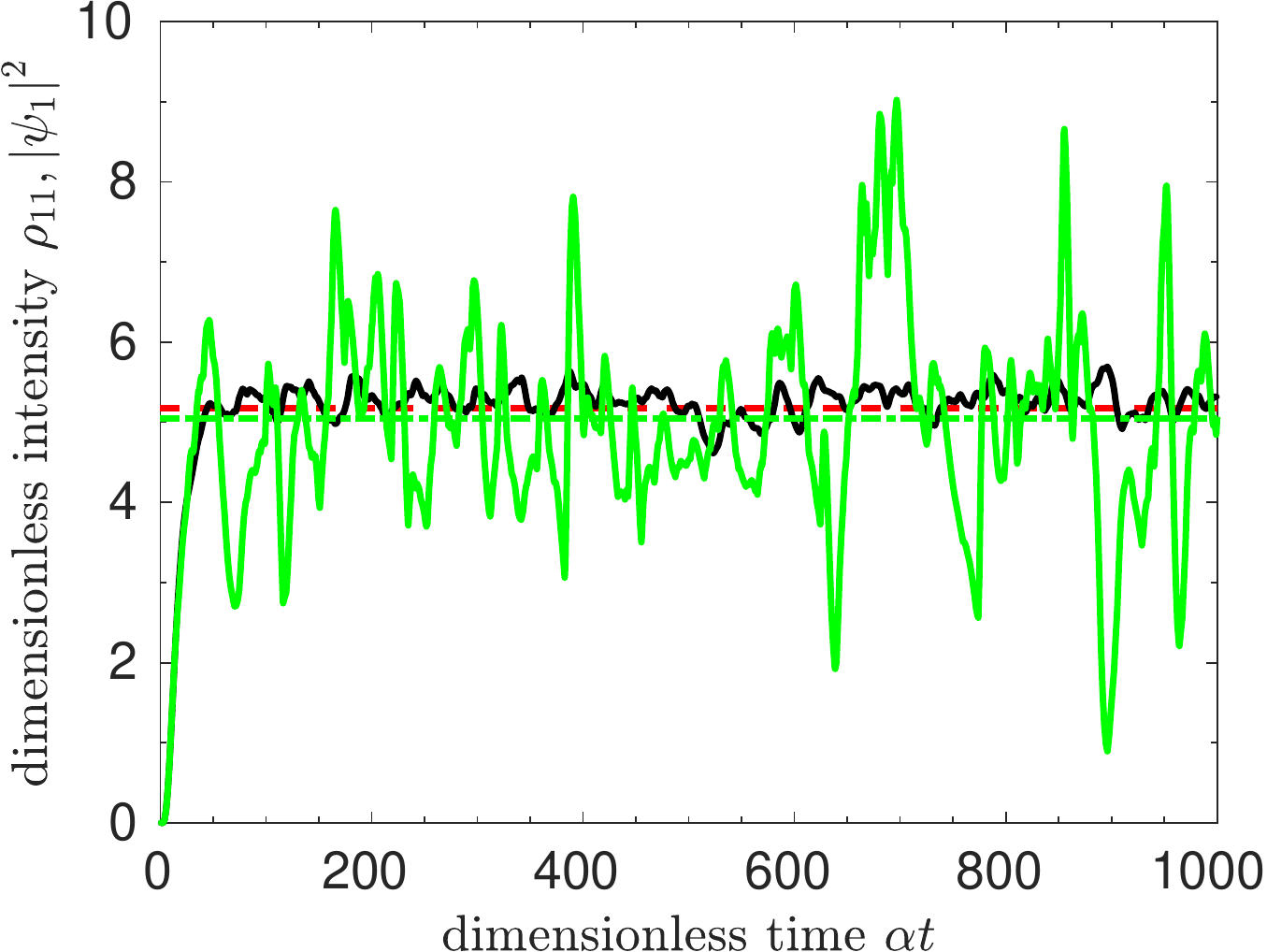}
\caption{Density matrix element $\rho_{11} = \ensembleaverage{|\psi_1|^2}$ and intensity $|\psi_1|^2$ vs (dimensionless) time $\alpha t$ for the \PT-symmetric dimer [Eqs.~(\ref{eq:e1H}), (\ref{eq:e1Hn}), and~(\ref{eq:eqLtot})]. The light green curve is the intensity $|\psi_1|^2$ belonging to an individual solution of the stochastic differential equation~(\ref{eq:ito}) with the time step $\alpha\Delta t = 0.001$ and the initial condition $\ket{\psi(0)} = 0$. Its long-time average after transient, $\alpha t\in [100,1\,000]$, is marked by the dash-dotted green line. The dark black curve is an ensemble average over 100 random realizations. The red dashed line marks $\rho_{11}$ of the stationary solution of Eq.~(\ref{eq:drhodt}). The physical parameters are $\pf/\alpha = 1 = \omega_0/\alpha$, $\varepsilon = 0$, $\noiseps/\alpha = 2.5\cdot 10^{-3}$, $\kappa = 1.5\kappac$, and $\ket{P}/\alpha = (0.1,0)^\transpose$.}
\label{fig:rho11}
\end{figure}

Figure~\ref{fig:rho11} provides a verification of the analytical results for the \PT-symmetric dimer against numerical simulations using the Financial Toolbox of MATLAB. Shown is the dynamics of the intensity (individual, ensemble-averaged, long-time averaged) in the first site of the dimer.
The damping rate $\kappa$ is chosen to be slightly larger than the critical rate~$\kappac$ in Eq.~(\ref{eq:e1kappac}) to ensure stability. Choosing $\kappa < \kappac$ leads to unstable dynamics (not shown).
It can be clearly seen that after a transient the ensemble average quickly settles down to the asymptotic stationary solution of Eq.~(\ref{eq:drhodt}). The individual realization exhibits strong fluctuations which is due to a rather large $\noiseps$ and due to the fact that for the chosen $\kappa$ the system is near the instability.
Nevertheless, the long-time average of the individual realization is very close to the stationary solution of Eq.~(\ref{eq:drhodt}). Hence, a single measurement averaged over time equals an ensemble average as it is expected here. An equally good agreement is observed for other parameter sets (not shown). 

\begin{figure}[ht]
\includegraphics[width=0.85\columnwidth]{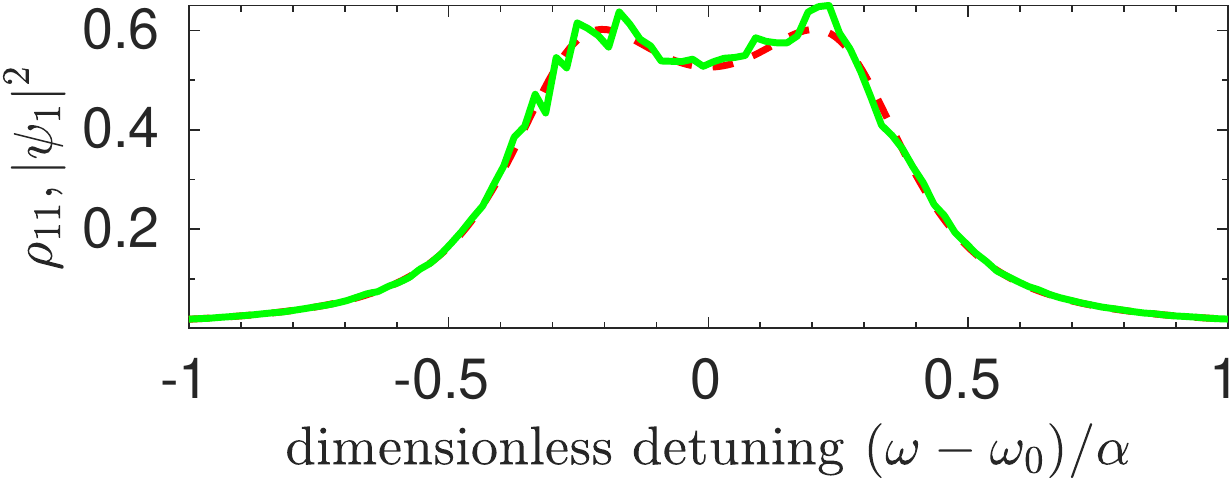}
\caption{Density matrix element $\rho_{11} = \ensembleaverage{|\psi_1|^2}$ and time-averaged intensity $|\psi_1|^2$ vs (dimensionless) detuning~$(\pf-\omega_0)/\alpha$ for the \PT-symmetric dimer [Eqs.~(\ref{eq:e1H}), (\ref{eq:e1Hn}), and~(\ref{eq:eqLtot})]. The light green curve is the long-time average of $|\psi_1|^2$ as a solution of the stochastic differential equation~(\ref{eq:ito}). The red dashed curve is $\rho_{11}$ belonging to the stationary solution of Eq.~(\ref{eq:drhodt}). The numerical and physical parameters are as in Fig.~\ref{fig:rho11} except that $\pf$ is varied on a uniform grid of 100 discretization points and the perturbation strength~$\varepsilon$ is set to $2\kappac^2/\alpha$ which guarantees a clear detection of the frequency splitting.}
\label{fig:spectrum}
\end{figure}

Results for finite perturbation strength~$\varepsilon$ and varying pump frequency~$\pf$ are presented in Fig.~\ref{fig:spectrum}. Very good agreement between the long-time average of individual solutions (one for each frequency) and the stationary solution of Eq.~(\ref{eq:drhodt}) is visible. The small deviations are due to the finite integration time.

\begin{figure}[ht]
\includegraphics[width=0.95\columnwidth]{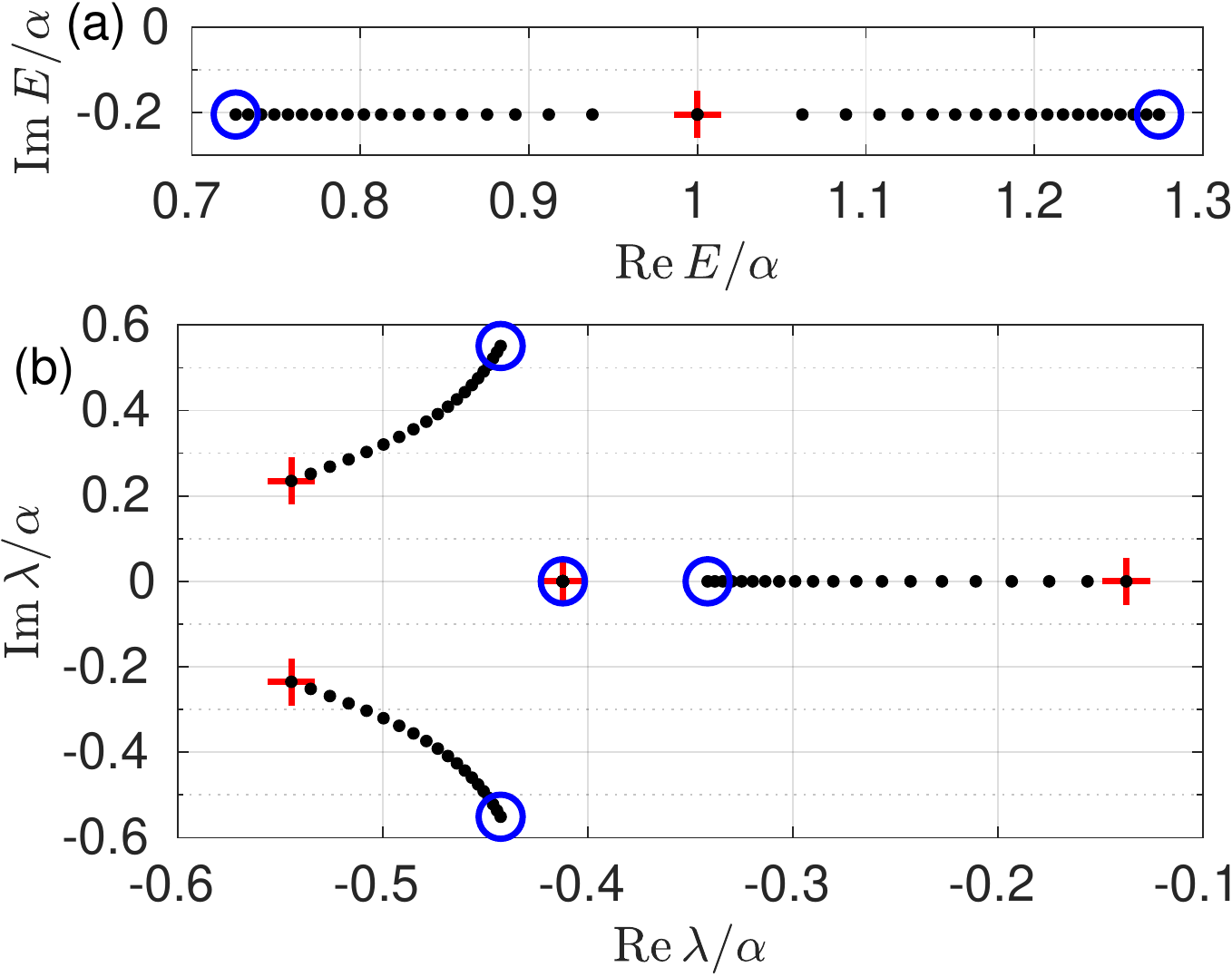}
\caption{Complex spectrum of (a) the Hamiltonian~$\Heff$ and (b) the Liouvillian~$\oL$ for the \PT-symmetric dimer. The former and latter are computed by (numerically) diagonalizing Eqs.~(\ref{eq:e1Heff}) and~(\ref{eq:eqLtot}). The physical parameters are as in Fig.~\ref{fig:rho11} except that the perturbation strength~$\varepsilon$ varies from $0$ [plus symbol(s)] to $2\kappac^2/\alpha$ (open circles) in 20 equidistant steps (dots).}
\label{fig:evLH}
\end{figure}

Figure~\ref{fig:evLH} compares the Hamiltonian spectrum with the Liouville spectrum. For $\varepsilon = 0$ and $\noiseps \neq 0$ the Hamiltonian~$\Heff$ is at an EP (plus symbol) but the Liouvillian~$\oL$ is not (four plus symbols). Increasing $\varepsilon$ leads to a splitting of the eigenvalues of the Hamiltonian. For the maximal $\varepsilon$ shown here the (real) splitting is large enough to be resolvable in a spectrum (cf. Fig.~\ref{fig:spectrum}). In Fig.~\ref{fig:evLH} the EP-enhanced splitting is visible by the larger separation of the dots near the EP. 
In the regime of large $\varepsilon$ where the effect of the noise on the frequency splitting can be ignored, the eigenvalues of~$\oL$ approach the eigenvalues of $\oL'$ in Eq.~(\ref{eq:Lindbladwqj}) which are $-i(E_k-E_l^*)$ with $E_k$ being the eigenvalues of~$\Heff$. Specifically, it means that the real part of the eigenvalue of~$\oL$ with the largest real part decreases with increasing $\varepsilon$ which further stabilizes the sensor.

\subsection{Microcavity sensor based on asymmetric backscattering}
\label{sec:example2}
As a second example we consider the setup in Ref.~\cite{COZ17} where an HEP$_2$ is realized in an optical microtoroid controlled by two nanofiber tips. The corresponding non-Hermitian Hamiltonian in the basis of counterclockwise and clockwise traveling waves is
\begin{equation}\label{eq:e2Hs}
\Hs = \left(\begin{array}{cc}
\Omega_0 & A_0\\
0   & \Omega_0\\
\end{array}\right) 
\end{equation}
with the complex frequency $\Omega_0$ and the complex off-diagonal element $A_0$. The latter describes backscattering of light from clockwise to counterclockwise traveling direction. The backscattering in the opposite direction is absent. This is an example of an HEP generated by fully asymmetric backscattering~\cite{WKH08,WES11,Wiersig11,POL16,Wiersig18b}. Note that in the chosen basis of traveling waves the diagonal elements of the Hamiltonian~(\ref{eq:e2Hs}) are equal due to reciprocity~\cite{Wiersig16}.
The same applies to the perturbation induced by a target particle, 
\begin{equation}
\Hp = \left(\begin{array}{cc}
\Omega_1 & A_1\\
B_1 & \Omega_1\\
\end{array}\right) \ .
\end{equation}

Fluctuations in the positions of the nanofiber tips can introduce noise into the system. The whispering-gallery cavity perturbed by two local perturbations at azimuthal position $\beta_j$ can be described by~\cite{Wiersig11}
\begin{equation}\label{eq:Hamiltonian}
\op{H}^{(2)} = \left(\begin{array}{cc}
\Omega^{(2)} & A^{(2)} \\
B^{(2)} & \Omega^{(2)}\\
\end{array}\right) 
\end{equation}
with 
\begin{eqnarray}
\Omega^{(2)} & = & \Omega^{(0)}+\sum_{j=1}^2(V_j+U_j) \ ,\\
\label{eq:AN}
A^{(2)} & = & \sum_{j=1}^2(V_j-U_j)e^{-i2m\beta_j} \ ,\\
\label{eq:BN}
B^{(2)} & = & \sum_{j=1}^2(V_j-U_j)e^{i2m\beta_j} \ .
\end{eqnarray}
The complex numbers $2V_j$ and $2U_j$ are frequency shifts for positive- and negative-parity modes introduced by local perturbation $j$ alone. The positive integer $m$ is the azimuthal mode number. Considering small angular fluctuations $d\beta_j = dW_j(t)$ with $j = 1,2$ in a small time interval $dt$, Taylor expanding Eqs.~(\ref{eq:Hamiltonian})-(\ref{eq:BN}) gives
\begin{equation}\label{eq:e2Hn}
\Hnoisej = \left(\begin{array}{cc}
0 & ia_j\\
-ib_j & 0\\
\end{array}\right) 
\end{equation}
with 
\begin{eqnarray}
a_j & = & -2m(V_j-U_j)e^{-i2m\beta_j}\ ,\\
b_j & = & -2m(V_j-U_j)e^{i2m\beta_j} \ .
\end{eqnarray}
Note that $b_j \neq a_j^*$ in general, but $|b_j| = |a_j|$. Each nanofiber tip is a source of independent noise, i.e., $\nsources = 2$. It is natural to assume that the strength of the angular fluctuations are equal, i.e., $\noiseps_1 = \noiseps_2 = \noiseps$. In contrast to the example in Sec.~\ref{sec:example1} the fluctuations appear here in the coupling strength of the modes.

Since $\Hnoisej$ is in general non-Hermitian, the fluctuations may act as gain on the wave function dynamics. This can be seen more concretely by noting that $\Hnoisej^2 = a_jb_j\ident$ and therefore 
\begin{equation}\label{eq:e2Heff}
\Heff = \left(\begin{array}{cc}
\Omega_0+\varepsilon\Omega_1-i\noiseps\frac{\sigma^2}{2} & A_0 + \varepsilon A_1\\
\varepsilon B_1   & \Omega_0+\varepsilon\Omega_1-i\noiseps\frac{\sigma^2}{2} \\
\end{array}\right) \ ,
\end{equation}
with the complex number $\sigma^2 = a_1b_1+a_2b_2$. The requirement that the eigenvalues of the effective Hamiltonian~(\ref{eq:e2Heff}) have negative imaginary part implies that $\Gamma_0 = -2\imagc{\Omega_0}$ is positive (as it must be true also for the special case $\varepsilon = 0 = \noiseps$). 
Like in the example in Sec.~\ref{sec:example1}, the noise described by the Hamiltonian~(\ref{eq:e2Hn}) does not move the system away from the HEP in the effective Hamiltonian~(\ref{eq:e2Heff}). It just shifts the eigenvalue by $-i\noiseps\sigma^2/2$. Interestingly, if $\real{\sigma^2} < 0$ this slightly improves the resolvability of the frequency splitting.  
The fact that in both examples the HEP is not moved in parameter space is not a coincidence. It is natural that the noise induces a uniform dephasing. However, counterexamples can be constructed.
 
A straightforward calculation for the Liouvillian 
\begin{equation}\label{eq:eqLtot2}
\oL = \oLs+\varepsilon\oLp+\noiseps\sum_j\oLnoisej 
\end{equation}
gives 
\begin{equation}\label{eq:e2Ls}
\oLs = \left(\begin{array}{cccc}
-\Gamma_0 & -iA_0     & iA_0^*     &  0      \\
0         & -\Gamma_0 & 0         & iA_0^*      \\
0         & 0         & -\Gamma_0 &-iA_0    \\
0         & 0         & 0         &  -\Gamma_0\\
\end{array}\right) 
\end{equation}
and
\begin{equation}\label{eq:e2Lp}
\oLp = \left(\begin{array}{cccc}
-\Gamma_1       & -iA_1   & iA_1^*   &  0      \\
-iB_1    & -\Gamma_1       & 0       & iA_1^*     \\
iB_1^*   & 0       & -\Gamma_1       & -iA_1    \\
0       & iB_1^*   & -iB_1   &  -\Gamma_1      \\
\end{array}\right) 
\end{equation}
with $\Gamma_1 = -2\imagc{\Omega_1}$ and
\begin{widetext}
\begin{equation}\label{eq:e2Lnoise}
\oLnoisej = \left(\begin{array}{cccc}
-\real{a_jb_j}    &  0                 &  0                 &   |a_j|^2   \\
 0                    & -\real{a_jb_j} & -a_j^*b_j      &   0     \\
 0                    & -a_jb_j^*      & -\real{a_jb_j} &   0     \\
 |b_j|^2         &  0                 &  0                 &  -\real{a_jb_j}     \\
\end{array}\right) \ .
\end{equation}
\end{widetext}
In the case with noise but no perturbation the eigenvalues can be calculated in lowest order in $\noiseps$ to be $\lambda_3 = -\Gamma_0-\noiseps\real{\sigma^2}$ and 
\begin{equation}
\lambda_l = -\Gamma_0-e^{i2\pi(l+1/2)/3}\left[2\noiseps\left(|a_1|^2 + |a_2|^2\right)|A_0|^2\right]^{1/3}
\end{equation}
with $l = 0,1,2$. For small $\noiseps$ the eigenvalue $\lambda_1$ has the largest real part
\begin{equation}
\realc{\lambda_1} = -\Gamma_0+\left[2\noiseps\left(|a_1|^2 + |a_2|^2\right)|A_0|^2\right]^{1/3} \ .
\end{equation}
The sensor becomes dynamically unstable if $\realc{\lambda_1} > 0$. This instability is again a noise-induced parametric instability and is not related to a possible non-Hermiticity of the noise Hamiltonian~(\ref{eq:e2Hn}). 
Stability is ensured if
\begin{equation}\label{eq:e2cond}
\noiseps \leq \frac{\Gamma_0^3}{2\left(|a_1|^2 + |a_2|^2\right)|A_0|^2} \ .
\end{equation}
If this condition is not fulfilled then one has to add $-i\kappa\ident$ to the system Hamiltonian (and hence $-2\kappa\ident$ to $\oLs$) with damping rate $\kappa > \kappac = \realc{\lambda_1}/2$. Since $\Gamma_0 > 0$, the stabilization has no dramatic consequences -- in strong contrast to the \PT-symmetric system in the previous section -- because the nonnoisy limit has already a finite decay rate.

For a passive device the backscattering strength~$|A_0|$ is bounded from above by $2\Gamma_0$~\cite{Wiersig16}. The best case for resolving the frequency splitting, $|A_0| = 2\Gamma_0$, is the worst case for the dynamical stability. Here, Eq.~(\ref{eq:e2cond}) simplifies to 
\begin{equation}
\noiseps \leq \frac{\Gamma_0}{8\left(|a_1|^2 + |a_2|^2\right)} \ .
\end{equation}

\section{Conclusion}
\label{sec:conclusion}
We have introduced a Lindblad-type formalism to study the influence of parametric white Gaussian noise on the performance of sensors based on exceptional points. It has been revealed that the resolvability of the sensor is determined by an effective Hamiltonian (the system Hamiltonian plus a term related to the noise sources) and that the dynamical stability is determined by the Liouvillian describing the time evolution of the density operator. We have determined the condition under which the position of the Hamiltonian exceptional point in parameter space is not changed by the noise. In this case the noise may only broaden the spectral lines which does not spoil the resolvability considerably.

We have discussed the Liouville spectrum and its degeneracies. In particular, we have shown that if the system Hamiltonian is at an exceptional point of order~$\order$ then the system Liouvillian is at an exceptional point of larger order $2\order-1$. If additionally one of the eigenvalues is close to the imaginary axis this can lead to an instability of the sensor with respect to noise. This undesired instability can be removed by adding dissipation which, however, reduces the resolvability. This reduction in resolvability can be significant in particular for parity-time-symmetric sensors.

These findings have been illustrated with two examples, a parity-symmetric dimer and a whispering-gallery microcavity sensor with asymmetric backscattering. In both cases the position of the Hamiltonian exceptional point is unchanged, but the position of the Liouvillian exceptional point is changed leading in the first example to an instability. The dissipation that is needed to compensate this instability is calculated.  

We believe that our study will help to assess the performance of exceptional point-based sensors and to understand the relation of non-Hermitian degeneracies in Hamiltonians and Liouvillians.

\acknowledgments 
Fruitful discussions with R.~El-Ganainy, A.~Eisfeld, F.~Minganti, F.~Nori, and J. Kullig are acknowledged.

%

\bibliographystyle{aipnum4-1}

\end{document}